
\input harvmac.tex
\vskip 2in
\Title{\vbox{\baselineskip12pt
\hbox to \hsize{\hfill ms 7122}
\hbox to \hsize{\hfill RU-92-63}}}
{\vbox{\centerline{ Curvature Singularity as the Vertex Operator}}}
\centerline{Dimitri Polyakov\footnote{$^\dagger$}
{polyakov@physics.rutgers.edu}}
\medskip
\centerline{\it Dept. of Physics and Astronomy}
\centerline{\it Rutgers University, Piscataway, NJ 08855}
\vskip .5in
\centerline {\bf Abstract}
\smallskip
The 2d theory of free massless bosons,as an example of CFT,is discussed on the
manifold with the point-like singularity in the curvature.
We argue that such the singularity can be represented as the
non-local vertex operator inserted in the theory.This representation
 makes it possible  to get all the correlation functions.
It is shown how the presence of the "needle" in the geometry of the surface
produces the breakdown of the conformal symmetry.
Examples of the correlators  are derived.
\Date{December 92}
\vfill\eject
\lref\bpz{Belavin,Polyakov,Zamolodchikov,Nucl.Phys. {\bf B241} (1984)
333-380.}
\lref\cardy{John L.Cardy,Nucl.Phys. {\bf B240} (1984) 514-532.}
\lref\alv{Orlando Alvarez,Nucl.Phys. {\bf B216} (1983) 125-184.}
\lref\dist{J.Distler and H.Kawai,Nucl.Physics. {\bf B321} (1989)
509-527.}
\lref\seib{N.Seiberg,Common Trends in Mathematics and Quantum Field
Theory (1990)}
It is usually implied,when constructing 2d CFT,that we consider our
theory  on some 2-d manifold with a smooth metric.Indeed,it
is known that for every such manifold there exists a reparametrization
which makes the metric to be conformally equivalent to the flat one:
\eqn\lowen{g_{\mu\nu}= \Omega(z,\bar{z})\eta_{\mu\nu},}
where $\eta_{\mu\nu}$ is the Euclidean or Minkowski space-time metric.
Since the correlation functions in CFT are invariant under the Weil
transformations,the problem can thus be reduced to the theory on the surface
of zero curvature,which is well-defined~\refs{\bpz}.
 The situation might be different,however,if we consider the CFT on
the 2d surface with a metric to have a  point-like singularity .
For instance,one of it's eigenvalues  can be zero or infinity at this
point while the second one is to remain non-singular.
 This  metric may not be conformally equivalent to the flat one,and
the theory cannot be reduced to that  developed by BPZ and needs
significant modifications.Namely,the vacuum in this theory is
not necessarily SL(2,C) invariant,but can rather be represented as
some vertex operator acting  on the SL(2,C) invariant zero state
of the usual CFT.
 This paper regards one concrete example of such the situation.
{}From now on,unless otherwise is stated,we will be considering the theory
of free massless bosons on the surface with the radially symmetrical metric:
\eqn\lowen{ds^2=r^{2\lambda}d\varphi^2+dr^2,}
where $0\leq r<\infty,
      -\pi\leq\varphi\leq\pi$
      are the parameters of the surface,
  $ \lambda\neq
                0;1$.
For example,if $\lambda>1$,
 this metric defines a surface with a needle at zero.
For simplicity,let's also assume $\lambda$ to be integer.
By means of the reparametrization:
\eqn\lowen{w=\varphi + {ir^{1-\lambda}\over{1-\lambda}},}
\eqn\lowen{\bar{w}=\varphi - {ir^{1-\lambda}\over{1-\lambda}}}
we get the metric to have the form (1) and note that in terms of coordinate
system $(w,\bar{w})$ we have the theory defined on a semicylinder,up to the
scale factor $\Omega$.
Then, by using another reparametrization:
$ z=e^w,\bar{z}=e^{\bar{w}},$
we bring the metric to the form:
%
\eqn\lowen{ds^2={1\over{z\bar{z}}}[{{1-\lambda}\over2i}ln{z\over\bar{z}}]^{2\lambda\over{1-\lambda}}dzd\bar{z}}

Then,by performing the Weil transformation,we can eliminate the factor
before $dzd\bar{z}$.It is easily checked that the set $(z,\bar{z})$ defines the
plane without the unit disc,and,thus by performing the above
reparametrisations and Weil transformation,we can try to express our original
theory in terms of the theory on the plane without unit disc with the
flat metric. To do this,we need to check how the partition function
transforms under the above operations.Let's denote

$e^\sigma={1\over{z\bar{z}}}[{{1-\lambda}\over2i}ln{z\over\bar{z}}]^{2\lambda\over{1-\lambda}}$,
and
   $ \hat g =\eta_{\mu\nu}$.
 The partition function is invariant under a reparametrization.Thus,in
terms of $z,\bar{z}$ parameters it looks like
\eqn\lowen{Z=\int D_{{e^\sigma}\hat g}[\phi]e^{-i\int
d^{2}z\partial\phi\bar{\partial}\phi}}
The measure $D_{{e^\sigma}\hat g}[\phi]$ is defined so that
\eqn\lowen{\int D_{{e^\sigma}\hat
g}[\phi]e^{{-{||\delta\phi\||}^2}_{{e^\sigma}\hat g}}=1,}
where ${{||\delta\phi||}^2}_{{e^\sigma}\hat g}=\int
d^{2}z{\sqrt{g}}\delta\phi\delta\phi$.Here we should always keep in mind that
the set of values the parameters $z,\bar{z}$ can take defines the plane
with cut off unit disc in  the $z,\bar{z}$ coordinate system.
Upon being Weil transformed,the action is unchanged,while the measure
is  multiplied by the Liouville-like factor~\refs{\dist, \seib}:
\eqn\lowen{D_{{e^\sigma}\hat g}[\phi]\rightarrow D_{\hat
g}[\phi]e^{{1\over24\pi}(S_{l}(\sigma,\hat g)+{S_{l}}^{(1)}(\sigma,\hat g))}}
Here $$S_{l}(\sigma,\hat g)=\int
d^{2}z{\sqrt{g}}({{1\over2}\partial\sigma\bar{\partial}\sigma+\hat
R\sigma+\mu\e^{\sigma}})$$ is the usual Liouville action while
${S_{l}}^{(1)}$ is the "boundary" Liouville term.
The nature of the last one is easily understood.The measure is not
invariant under the Weil transformation due to the non-invariance of the
norm ${||\delta\phi||}^2=\int d^{2}z{\sqrt{g}}\delta\phi\delta\phi$-
this is where the regular Liouville term always comes from.However,
here this integral is  taken over  the region outside the unit
disc and therefore it depends on what boundary conditions we impose on the
circle.This gives rise to the additional factor  appearing in the
\vfill\eject
transformation law for the measure.This factor has been calculated
by Alvarez~\refs{\alv}, and it is given by:
\eqn\lowen{{S_{l}}^{(1)}(\sigma,\hat
g)=\int_{\partial{M}}ds(\sigma(s)+\lambda_{0}e^{{1\over2}\sigma(s)})}
in case if the boundary is a unit circle,
 where $\partial{M}
$ denotes the unit circle,s is the variable of
integration over it.
Thus,under the Weil transformation,the partition function is
transformed to:
\eqn\lowen{\int D_{{e^{\sigma}}{\hat g} }[\phi]{e^{-\int_{M}
d^{2}z\partial\phi\bar{\partial}\phi}}\rightarrow [\int
D_{{\hat g} }[\phi]{e^{-\int_{M}
d^{2}z\partial\phi\bar{\partial}\phi}}]e^{{1\over24\pi}(S_{l}(\sigma,\hat
g)+{S_{l}^{(1)}}(\sigma,\hat g))}}
Here M is the region outside the unit circle.
All the changes come from the non-invariance of the
${||\delta\phi||}^{2}
                     $ on M and $ \partial{M}$.
Note that the field should take a constant value on the unit circle
since in our theory the circle is the image of the singular point
obtained after the reparametrization and the Weil transformation.
 (The classical field equation in the above metric with
the radial symmetry has single valued solutions).

What we would like to do is to express the  theory of free
massless bosons  outside the
unit circle with some boundary conditions in terms of some other field
theory, which is not necessarily Weil-invariant  on the classical
level,but which is rather defined on the entire plane.
Our hope is that the theory
we can then obtain would be the usual non-singular theory of free
bosons with some inserted operator comprising all the
conformal non-invariance of the problem.
 To do this,let's notice that the classical Dirichlet's problem
(this is exactly the kind of problem we deal with since we know that the
field must be a constant on the boundary):
\eqn\lowen{\partial\bar{\partial}\phi=0;
  \phi|_{\partial{M}}=-i\phi_{0}}
has the solution
$ \phi=-i\int_{\partial{M}}ds{{\partial{G}}\over{\partial{\vec
n}}}\phi_{0}$,
where G is the Green's function of the Dirichlet's problem.
We can write down this expression as
\eqn\lowen{-i\int_{\partial{M}}ds{\partial_{\vec n}}G\phi_{0}=i\int
d^{2}z({\partial_{\vec n}}(\phi_{0}\delta_{s}))G.}
 The classical equation of motion with the boundary conditions (11)
is equivalent to the problem on the whole plane with the
corresponding term on the rhs:
\eqn\lowen{\partial\bar{\partial}\phi=i{\partial_{\vec
n}}(\phi_{0}\delta_{s})}
This classical equation of motion is given by the following
effective action:
\eqn\lowen{S_{eff}=\int
d^{2}z[\partial\phi\bar{\partial}\phi+i\phi{\partial_{\vec
n}}(\phi_{0}\delta_{s})]}
\vfill\eject
When quantizing the Dirichlet's problem we can keep doing the functional
integral with $S_{eff}$ with the field theory defined on the whole
plane,instead of doing it with the weight $S=\int
d^{2}z\partial\phi\bar{\partial}\phi$ with field configurations outside the
unit disc only.
 This is  nothing bizarre in this procedure since it can be
shown that the additional operator in the action  just
 screens off the field configurations inside the circle,so,nothing
principally new is added to the theory.This is somewhat similar
to the method of electrostatic images which was used by Cardy~\refs{\cardy}
for solving the CFT on a half-plane.
 So,the partition function of the theory can be written in terms of
  the free bosonic theory on the plane by introducing the following extra
term into the action:
\eqn\lowen{Z=<0_{sing}|0_{sing}>=e^{S_{l}+{S_{l}}^{(1)}}\int D_{\hat
g}[\phi]{e^{-\int d^{2}z[\partial\phi\bar{\partial}\phi+i\phi{\partial_{\vec
n}}{(\phi_{0}\delta_{s})}]}},}
where $|0_{sing}>,<0_{sing}|$ are the vacua of the singular theory.
Since $\int d^{2}z\phi{\partial_{\vec n}}(\phi_{0}\delta_{s})=-\oint ds
(\phi_{0}{\partial_{\vec n}}\phi)$,
we can modify this expression as
\eqn\lowen{Z=<0_{sing}|0_{sing}>=e^{S_{l}+{S_{l}}^{(1)}}<0|e^{i\oint
ds\phi_{0}{\partial_{\vec n}}\phi}|0>,}
where $|0>,<0|$ are just the well defined vacua of the theory of free
massless bosons.
The Liouville-like factor here is insignificant once we're going to
consider  the correlation functions only.
Indeed,since $<\phi_{1}...\phi_{N}>={\int
D[\phi]\phi_{1}...\phi_{N}e^{-S}}{1\over{\int D[\phi]e^{-S}}}$,
and since this factor is $\phi$-invariant,it will be cancelled when
we divide over Z.
 Therefore,the most meaningful correction the singularity contributes to
the theory is the boundary extra term in the action,which can as well be
thought as some inserted operator.In other words,we have shown that
the curvature singularity is represented by the non-local vertex
operator acting on the SL(2,C) invariant vacuum $|0>$:
\eqn\lowen{|0_{sing}>=e^{i\phi_{0}{\oint_{C_{0}}}ds{\partial_{\vec
n}}\phi}|0>,}
\eqn\lowen{<0_{sing}|=<0|}
Here $C_{0}$ is the contour around zero.
This inserted operator comprises the non-chirality the singular point
brings to the theory.Apart from this operator we may treat our theory
just as that of chiral and anti-chiral bosons on the plane with the flat
metric.
 We are now in the position to start computing the correlation functions
in our theory.First of all,let's calculate the simplest one-point
function - the average of the exponential operator
$:e^{\alpha\phi(z)}:$.
At the non-singular theory this function apparently should be zero-
it is the consequence of the conformal invariance.In our theory the vacuum
is no more conformally invariant since the
\vfill\eject
 the singular
point "recharges" the zero-state:
$|0>{\rightarrow}sing\rightarrow|0_{sing}>$.
We shall see that in our theory the one-point function is not
necessarily zero.
Thus
\eqn\lowen{<0_{sing}|:e^{\alpha\phi(z)}:|0_{sing}>
=<0|:e^{\alpha\phi(z)}::e^{-{\phi|_{C_{0}}}{\oint_{C_{0}}}
d\varphi{\partial_{\vec n}}\phi}:|0>.}
To compute this correlator let's figure out what is
$e^{{\phi|_{C_{0}}}{\oint_{C_{0}}}d\varphi{\partial_{\vec n}\phi}}$ in
terms of the mode expansion of  free massless bosonic field.
 We have:
$\phi(z,\bar{z})=\phi(z)+\bar{\phi}(\bar{z})$,
\eqn\lowen{i\partial\phi(z)=\sum_{n}{{\alpha_{n}}\over{z^{n+1}}},
\alpha_{n}={1\over2{\pi}i}\oint dz{z^n}i\partial\phi(z),}
and the same is for $\bar{\phi}(\bar{z})$ and $\bar{\alpha_{n}}$.
Also,
$\phi(z)=-i\phi_{0}-iln(z)\alpha_{0}+i\sum_{n\neq
0}{{\alpha_{n}}\over{n{z}^n}}$,
$-i\phi_{0}={1\over2{\pi}i}\oint dz{\phi(z)\over{z}}$,
and the similar things are true for ${\bar{\phi}}(\bar{z})$,
$\bar{\phi_{0}}$.
Thus
\eqn\lowen{\oint d\varphi{\partial_{\vec n}}\phi=-i\oint
dz\partial\phi(z)+i\oint
d\bar{z}\bar{\partial}\bar{\phi}(\bar{z})=-2{\pi}i(\alpha_{0}-\bar{\alpha_{0}})}
-on $C_{0}$.
Then, let's find out out what is $\phi$ on $C_{0}$ in terms of modes.
Since it is fixed to be a constant,the only  modes that should get through
are $\varphi$-independent.The modes that comply with this
restriction are $\phi_{0}$ and $\bar{\phi_{0}}$.Thus we have found
 the singularity operator defined on the unit circle:
\eqn\lowen{:e^{-{\phi|_{C_{0}}}{\oint_{C_{0}}}d\varphi{\partial_{\vec
n}}\phi}:=:e^{2{\pi}(\phi_{0}-{\bar{\phi_{0}}})(\alpha_{0}-{\bar{\alpha_{0}}})}:.}
Therefore,expanding the exponents,we get:
\eqn\grav{\eqalign{<0_{sing}|:e^{\alpha\phi(z)}:|0_{sing}>=\sum_{n,m=0}^\infty{{\alpha^{n}{2\pi}^{m}}\over{n!m!}}<0|:{\phi(z)}^{n}::{(\phi_{0}-\bar{\phi_{0}})}^{m}{(\alpha_{0}-\bar{\alpha_{0}})}^{m}:|0>=\cr
\sum_{k,l,m,n=0}^\infty{{\alpha^{n}{2\pi}^{m}{(-1)}^{2m-k-l}}\over{n!m!}}{C_{m}}^{k}{C_{m}}^{l}<0|:{\phi(z)}^{n}::{\phi_{0}}^{k}{\bar{\phi_{0}}}^{m-k}{\alpha_{0}}^{l}{\bar{\alpha_{0}}}^{m-l}:|0>=\cr
\sum_{k,l,m,n=0}^\infty{{\alpha^{n}{2\pi}^{-m}{(-1)}^{2m-k-l}}\over{n!m!}}{C_{m}^{k}{C_{m}}^{l}\oint
...\oint
{{dz_{1}}\over{z_{1}}}...{{dz_{k}}\over{z_{k}}}{{d\bar{z_{k+1}}}\over\bar{z_{k+1}}}...{{d\bar{z_{m}}}\over\bar{z_{m}}}\cr
{dw_{1}...dw_{l}d\bar{w_{l+1}}...d\bar{w_{m}}}}\cr
(<0|:{\phi(z)}^{n}::\phi(z_{1})...\phi(z_{k})\bar{\phi}(\bar{z_{k+1}})...\bar{\phi}(\bar{z_{m}})\partial\phi(w_{1})...\partial\phi(w_{l})\bar{\partial}\bar{\phi}(\bar{w_{l+1}})...\bar{{\partial}\bar{\phi}(\bar{w_{m}}):|0>)}}}
Thus we have expressed this correlator in terms of free
fields.Therefore,the Wick's theorem
\vfill\eject
is now appliable.
First,it's clear that no terms with contractions like
$\phi(z)\longleftrightarrow\bar{\phi}(\bar{z})$ and
$\phi(z_{i})\longleftrightarrow\bar{\partial}\bar{\phi}(\bar{w_{j}})$ would
contribute to the correlator.
Also,the "most contractible" case only is relevant,i.e. the sum will be
contributed by the  terms where all the operators are involved in
the contraction.Otherwise,we would get,up to a factor,some normal
ordered operator acting on the "flat" vacuum,which would give
us zero.All the said above yields the following constraints:
$1)k=m$;$2)l=m$;$3)n=2m$.The last statement implies that even n's only
are admissible.Hence,the one-point correlator in the singular theory
becomes:
%
\eqn\grav{\eqalign{<0_{sing}|:e^{\alpha\phi(z)}:|0_{sing}>=\sum_{m=o}^\infty{{\alpha}^{2m}\over{{2\pi}^{m}m!(2m!)}}\oint...\oint
{{dz_{1}}\over{z_{1}}}...{{dz_{m}}\over{z_{m}}}dw_{1}...dw_{m}\cr
[\prod_{i=1}^{m}<0|\phi(z)\phi(z_{i})|0>\prod_{j=1}^{m}<0|\phi(z)\partial\phi(w_{j})|0>(2m!)]=\cr
=\sum_{m=0}^\infty{{\alpha}^{2m}\over{{(2\pi)}^{m}m!}}\oint...\oint
{{dz_{1}}\over{z_{1}}}...{{dz_{m}}\over{z_{m}}}dw_{1}...dw_{m}[\prod_{i=1}^{m}ln(z-z_{i})\prod_{j=1}^{m}{1\over{z-w_{j}}}]=\cr
=\sum_{m=0}^\infty{{(-1)}^{m}\over{m!}}{[2\pi{\alpha}^{2}ln(z)]}^{m}=e^{-2\pi{\alpha}^{2}ln(z)}={1\over{z^{2\pi{\alpha}^{2}}}}.}}
This means we have the critical exponent for how the singularity impairs
the conformal symmetry of the theory.Since the CFT describes the systems
at the second order phase transition point,and we have the infinite
radius of correlations at this point,the above result confirms our
expectation of the long-range impairment of the conformal symmetry produced by
the singular point.
 The second thing we would like to calculate is the correlation function
between two singularities themselfes,describing the interaction of two
needles in the massless bosonic theory.Provided these points are
sufficiently apart from each other,we again may represent them as
two non-local vertex operators (the requirement for their being apart
comes from that the disturbances they contribute to the geometry
should not overlap).For simplicity,let's suppose that the first "needle"
is at zero,while the second one is on the real axis at the distance R
from the first.Let's again figure out what is the second operator in
terms of modes.The procedure is exactly the same as for the first one,
just the contour $C_{0}$ should be replaced by $C_{1}$-the unit circle
around the point $x=R$.Then,proceding exactly as above,we find:
\vfill\eject
\eqn\grav{\eqalign{-{\phi|_{C_{1}}}\oint d\varphi{\partial_{\vec
n}}\phi=-\oint_{C_{1}}dz\sum
{{\alpha_{n}}\over{z^{n+1}}}-\oint_{C_{1}}d\bar{z}\sum
{{\bar{\alpha_{n}}}\over{\bar{z^{n+1}}}}=0 }},
since apparently there are no poles if $R>1$.This means that the second
singularity operator in the correlation function is just the identity
operator.Therefore,this correlation function is just the constant:
\eqn\lowen{<0|:e^{-\phi|_{C_{0}}\oint_{C_{0}} d\varphi\partial_{\vec
n}\phi}::e^{-\phi|_{C_{1}}\oint_{C_{1}}d\varphi\partial_{\vec
n}\phi}:|0>=<0|:e^{-\phi|_{C_{0}}\oint_{
C_{0}}d\varphi\partial_{\vec
n}\phi}::1:|0>={Z\over{Z}}=1.}
This result is quite natural if we keep in mind how the singularity operator
acts: it screens off the field incide the unit circle while
the field outside is unchanged upon being acted by this operator.
Since nothing outside the circles surrounding the singular
points is affected,the interaction between two singularities,i.e.
their  correlation function must not depend on the distance between them
unless the circles of the correspondent operators cross each other.
On the contrary,if we formally make two circles to intersect,we would
immediately get the simple poles for the integrals in (25) and the
expression for the correlator in this case would be complicated,
in accordance with our expectations.However,we should stress that
this procedure would not be  lawful since we may only
consider the operator representation for two singularities if $R>>1$ and
the formal result for $R<1$ would hardly imply any physics.
However,we should emphasize that this result is only correct in our
very rough  approximation allowing us to consider two
operator representations independently.Strictly speaking,there should be
the interaction between singularities in more accurate approximations,
yet this question is not discussed in this paper.
Using the formalism developed above,we can in principle get all the
correlation functions in our theory,though technically it may require
long calculations.
However,we will be more interested to
compute $:T(z):|0_{sing}>$,which should give us the vacuum energy shift
provided by the singularity,i.e. a kind of the "Casimir effect".
\eqn\grav{\eqalign{:T(z):|0_{sing}>=:\partial\phi(z)\partial\phi(z)::e^{2\pi(\phi_{0}-\bar{\phi_{0}})(\alpha_{0}-\bar{\alpha_{0}})}:|0>=\cr
=\sum_{k,l,m}{{2\pi}^m\over{m!}}{C_{m}}^{k}{C_{m}}^{l}\oint...\oint{{dz_{1}}\over{z_{1}}}...{{dz_{k}}\over{z_{k}}}{{d\bar{z_{k+1}}}\over{\bar{z_{k+1}}}}...{{d\bar{z_{m}}}\over{\bar{z_{m}}}}dw_{1}...dw_{l}d\bar{w_{l+1}}...d\bar{w_{m}}\cr
(:\partial\phi(z)\partial\phi(z)::\phi(z_{1})...\phi(z_{k})\bar{\phi}(\bar{z_{k+1}})...\bar{\phi}(\bar{z_{m}})\partial\phi(w_{1})...\partial\phi(w_{l})\bar{\partial}\bar{\phi}(\bar{w_{l+1}})...\bar{\partial}\bar{\phi}(\bar{w_{m}}):|0>).}}
 Again,now we can proceed with doing Wick's contractions,according to the
 method developed. We are mostly interested in the  term in the
expansion which is proportional to $1\over{z^{2}}$.This corresponds to the
case when the both of $\partial\phi's$ in the stress-energy tensor participate
 in the contraction.The simple calculation gives us:
$:T(z):|0_{sing}> =
-{64{{\pi}^{4}}\over{z^{2}}}{:{(\alpha_{0}-\bar{\alpha_{0)}}}^{2}:}|0_{sing}>+$less
singular terms.
The operator which is multiplied by $1\over{z^{2}}$ in the OPE should
 give us the eigenvalues of $L_{0}$ acting on the singular
vacuum.Our aim now is to express them in terms of the eigenvalues
of $\alpha_{0}$ and $\bar{\alpha_{0}}$ acting on the regular vacuum.
We get:
\eqn\grav{\eqalign{:T(z):|0_{sing}>={{-64{\pi}^{4}}\over{z^2}}{(\alpha_{0}-\bar{\alpha_{0}})}^{2}e^{2\pi(\phi_{0}-\bar{\phi_{0}})(\alpha_{0}-\bar{\alpha_{0}})}|0>=\cr
{{-64{\pi}^{4}}\over{z^2}}\sum
{{(2\pi)}^{m}\over{m!}}{(\alpha_{0}-\bar{\alpha_{0}})}^{2}{(\phi_{0}-\bar{\phi_{0}})}^{m}{(\alpha_{0}-\bar{\alpha_{0}})}^{m}|0>.}}
Here we should make use of
$[\alpha_{0},\alpha_{0}]=[\alpha_{0},\bar{\alpha_{0}}]=[\bar{\alpha_{0}},\phi_{0}]=[\alpha_{0},\bar{\phi_{0}}]=[\phi_{0},\phi_{0}]=[\bar{\phi_{0}},\bar{\phi_{0}}]=[\phi_{0},\bar{\phi_{0}}]=0$,and
$[\alpha_{0},\phi_{0}]=[\bar{\alpha_{0}},\bar{\phi_{0}}]={1\over{2\pi}}$.
These relations are to be checked directly,by using the definitions (20)
and
$[\phi(z),\partial\phi(w)]=i{\delta}^{(2)}(z-w)$,$[\bar{\phi}(\bar{z}),\bar{\partial}\bar{\phi}(\bar{w})]=-i{\delta}^{(2)}(\bar{z}-\bar{w})$.
Therefore,by commuting the operators,we get:
\eqn\grav{\eqalign{:T(z):|0_{sing}>=-{{64\pi}\over{z^2}}\sum_{m}{{(2\pi)}^{m}\over{m!}}{(\alpha_{0}-\bar{\alpha_{0}})}^{2}{(\phi_{0}-\bar{\phi_{0}})}^{m}{(\alpha_{0}-\bar{\alpha_{0}})}^{m}|0>=\cr
=-{{64\pi}\over{z^2}}\sum_{m}{{(2\pi)}^{m}\over{m!}}:[{(\phi_{0}-\bar{\phi_{0}})}^{m}{(\alpha_{0}-\bar{\alpha_{0}})}^{m+2}+{2m\over{\pi}}{(\phi_{0}-\bar{\phi_{0}})}^{m-1}{(\alpha_{0}-\bar{\alpha_{0}})}^{m+1}+\cr
+{{m(m-1)}\over{\pi^2}}{(\phi_{0}-\bar{\phi_{0}})}^{m-2}{(\alpha_{0}-\bar{\alpha_{0}})}^{m}]:|0>.}}
Now,let $P_{0},\bar{P_{0}}$ be the eigenvalues of operators
$\alpha_{0},\bar{\alpha_{0}}$ acting on the vacuum of the "flat" theory.
They correspond to different values of the center of mass momentum.
Then,performing the summations in (29) we get the answer:
\eqn\lowen{:T(z):|0_{sing}>=-{{64\times9{\pi}^{4}}\over{z^2}}{(P_{0}-\bar{P_{0}})}^{2}|0_{sing}>+...}
This result might be interpreted as the appearance of some "effective
mass" in the theory due to the conformal symmetry breaking caused by
the curvature singularity at zero.
\vskip .5in
\centerline {\bf Discussion}
\bigskip
We have shown that the singular behavior of the 2d metric's eigenvalues
makes the vacuum to be conformally non-invariant.Namely,the metric
singularity is equivalent to the insertion of the non-local operator to
the Hilbert space of the theory.The singular field theory is quite equivalent
to the field theory with the boundary-this is what makes the theory
 easily solvable in the case of free massless bosons.
However,in our approach to this problem we crucially used the fact that
the original theory could be described
\vfill\eject
by the Lagrangian.This approach
doesn't work if we consider the field theory on a singular manifold which
does not have any classical equation of motion - for instance,it could
be some non-unitary minimal model in the CFT.Although it seems to be
plausible that the conformal symmetry would be again broken by the
"needle" and we somehow can again consider the operator representation,
it's unclear how to get the corresponding vertex operator
in  case when we have an operator algebra instead of the Hamiltonian.
Another interesting question would be to develop the similar formalism in the
string theory with a singular background.

\centerline {\bf Acknowledgements}
This work was supported by the Graduate Research Assistantship of
Rutgers University.I feel especial gratitude to V.Brazhnikov,V.Gurarie,
M.O'Loughlin and M.Zyskin for numerous discussions.


\vskip .5in
\footatend\vfill\immediate\closeout\rfile\writestoppt
\centerline{\bf{ References}}\bigskip{\frenchspacing%
\parindent=20pt\escapechar=` \input refs.tmp\vfill\eject}\nonfrenchspacing

\end